\documentstyle[12pt,psfig]{article}
\begin{document}

\input epsf

\vskip 1truecm
\rightline{Preprint  MCGILL-97/23}
\rightline{ e-Print Archive: hep-lat/9709053}
\vspace{0.2in}
\centerline{\Large $O(a)$ errors in 3-D SU($N$) Higgs theories}
\vspace{0.3in}

\centerline{{\Large Guy D. Moore}\footnote{e-mail address:  
	guymoore@hep.physics.mcgill.ca}}

\medskip

\centerline{\it Physics department, McGill University}
\centerline{\it 3600 University St.}
\centerline{\it Montreal, Quebec H3A 2T8 Canada}

\vspace{0.2in}

\centerline{\bf Abstract}

We compute the matching conditions between lattice and 
continuum 3-D SU($N$) Higgs theories, with both adjoint and 
fundamental scalars, at $O(a)$, except for additive corrections to
masses and $\phi^2$ insertions.  

\begin{verse} 
PACS numbers:  11.15.Ha, 11.15.Kc
\end{verse}

\section{Introduction}
\label{introduction}

The study of early universe cosmology demands the use of particle
physics, because Hubble expansion means that the universe began at a
very large temperature and density.  By looking back to time scales on
order seconds and temperatures on order 1MeV, and using QED, nuclear
physics, low energy weak interaction physics, and nonequilibrium
kinetics, it has been possible to predict the primordial abundances of
elements.  It is natural to try to push this success back to shorter
times and higher energies, where one finds a sequence of interesting
phase transitions:  the QCD phase transition, which it may soon be
possible to study through heavy ion collisions; the electroweak phase
transition, which may be responsible for baryogenesis; and perhaps even
GUT phase transitions, which may have bearing on baryogenesis,
inflation, and vacuum stability.

Our attempts to understand these phase transitions, and thermal physics
at these energy scales more generally, have been hampered
for two reasons.  First is an uncertainty about the underlying physics,
in the electroweak case and especially in the GUT case.  The second is
the infrared problem of interacting, bosonic thermal field theories.
Even at weak coupling, if one is interested in sufficiently long
wavelength phenomena, mutual interactions between light bosonic species
cause ordinary perturbation theory to break down.  Perturbation theory
is inevitably useless in the high temperature phase of Yang-Mills Higgs
theories when one considers length scales of order $1/g^2 T$, and for
shorter length scales this problem invariably enters calculations at
some perturbative order, because of interactions with modes at this
scale. 

This problem was first noted by Linde \cite{Linde}, and has since been
understood as arising from the essentially 3 dimensional nature of the
thermodynamics of the infrared, bosonic fields.  The partition function
describing equal time, equilibrium thermodynamics, $Z = {\rm Tr} e^{-
\beta H}$, can be written as a Euclidean path integral where ``time'' is
periodic with period $\beta = 1/T$.  At weak coupling, the
decoupling theorem allows one to integrate out Fourier modes with
nonzero wave number in this direction, the nonzero Matsubara frequency
modes\cite{oldDR}.  This leaves a three dimensional theory, which is
superrenormalizable, and therefore well behaved in the ultraviolet but
potentially poorly behaved in the infrared.  The couplings of this
theory have nonzero engineering dimension, which sets a natural scale 
where the perturbative expansion becomes unreliable.  The physics of 
interest often lies in this region.

The solution to this problem is to perform the integration over the
nonzero Matsubara frequencies analytically, reducing the theory to the
simplest sub-theory which contains the correct problematic infrared
behavior.  Then one studies the 3-D theory nonperturbatively, for
instance on a lattice.  The lattice treatment also turns out to be
particularly tractable because of the reduced dimension and the
superrenormalizability.  

In the case of the electroweak phase transition this program has been
pushed nearly to completion.  The dimensional reduction has been
performed \cite{FKRS1}, and the match between the 3-D continuum theory
and a 3-D lattice theory has been carried out so that they will have the
same small lattice spacing $a$ limit \cite{FKRS,Laine}.  This matching
procedure can be exact because of the superrenormalizability of the 3-D
theory.  Several groups have studied the resulting system numerically 
\cite{KLRSresults,Kripfganz,Teper,MooreTurok,KLRSSU2U1}.  
There has also been a very general analysis of
the dimensional reduction step \cite{KLRS}, which has been applied to
the supersymmetric extension of the theory as well \cite{MSSMDR}.

Numerical studies of the lattice theory display quite strong $O(a)$ 
systematic errors, which at first were removed by extrapolation 
\cite{KLRSresults}.  Except for an additive correction to $\Phi^2$ and 
the value of the Higgs mass parameter at the equilibrium point, all 
of these errors arise from one loop corrections to the lattice 
continuum match and can be absorbed into a renormalization of the 
couplings and wave functions of the lattice theory.  In particular 
such a renormalization removes all $O(a)$ errors in the determination 
of observables which describe 
the strength of the phase transition.  The details of this $O(a)$
improvement step, in the case of SU(2)$\times$U(1) Higgs theory, are
worked out in \cite{Oapaper}.  (There is a numerically unimportant
combinatorial error there, and two in Appendix C.  We will correct them
here.)  

Recently there has been interest in applying the dimensional reduction
program to several other theories.  It may be possible to understand the
thermodynamics of hot QCD above 
the chiral phase transition in terms of SU(3) + adjoint Higgs theory
\cite{KLRSadjoint,mDresults}.  Superconductivity is described by a
Landau-Ginsburg model which is 3-D U(1) + Higgs theory, which also
describes the Kibble mechanism for cosmic string formation.  It has
been the topic of recent study \cite{U1study}.  If the stop squark is
light, then it may be necessary to study the supersymmetric Standard
Model by dimensional reduction to SU(3)$\times$SU(2) theory with a
fundamental Higgs field for each group \cite{Lainestop}.  It will also be
necessary to use dimensional reduction and a numerical study of the
reduced theory to understand the details of GUT phase transitions, such
as the SU(5) breaking phase transition \cite{Rajantie}.

In each case it is best to perform both the dimensional reduction
calculation and the lattice to continuum matching calculation before
undertaking extensive numerical work.  In most cases the dimensional
reduction step has been performed; otherwise the generic rules derived in
\cite{KLRS} can be used.  Recently, Laine and Rajantie have computed the
lattice continuum relations for scalar masses and the order parameter
$\Phi^2$, for 3-D SU($N$) gauge theory with either a fundamental or an
adjoint Higgs field, at two loops \cite{LaineRajantie}.  
This is sufficient to ensure the
lattice model has the correct small $a$ limit.  It seems
profitable also to extend the $O(a)$ matching between the lattice and
continuum theories to the general case, so that measurables relating to
the strength of the phase transition can be computed up to $O(a^2)$
corrections.  We do this here.

Here is an outline of the remainder of the paper.  In Section \ref{idea}
we will explain the general idea of the $O(a)$
improvement, and why it is possible.  Section \ref{details} will present
the coefficients of the $O(a)$ corrections for SU($N$) + adjoint Higgs +
fundamental Higgs, with comments on how to remove either Higgs field.
Section \ref{extension} shows what to do when the gauge group is
SU($N$)$\times$SU($M$), each with Higgs fields.  The last section is the
conclusion.  There are also two appendicies; the first estimates
$O(a^2)$ errors by computing a particularly simple and important subset
of graphs, and the second gives numerical values in particular theories
and step by step instructions for building $O(a)$ improved lattice
actions. 

\section{$O(a)$ improvement:  the general idea}
\label{idea}

Considerable attention has been given to the problem of improving the
match between lattice and continuum regulated quantum field theories.
Most attention has focused on the case of 4-D theories, which are
relevant for vacuum field theory.  For instance, in 4-D pure SU($N$)
Yang-Mills theory with no fermions, the continuum action is
\begin{equation}
\int d^4 x \frac{1}{2} {\rm Tr} F_{\mu \nu} F_{\mu \nu} \, ,
\end{equation}
where $F_{\mu \nu} = F_{\mu \nu}^a T^a = (i/g_c) [ D_\mu , D_\nu ]$, and
$D_\mu = \partial_\mu - i g_c T^a A^a_\mu$ is the the fundamental
representation covariant derivative.  Here it is understood that
divergences are to be removed in the $\overline{\rm MS}$ scheme at a
renormalization point $\mu$ and that $g_c$ refers to the value at that
renormalization point.

A corresponding lattice action is
\begin{equation}
\frac{2}{g_l^2 a^4} a^4 \sum_{x} \sum_{i < j} {\rm Tr} \left( {\bf 1} -
	P_{ij}(x) \right) \, ,
\end{equation}
where $P_{ij}(x)$ is the product of link matricies around a square in the
$i,j$ direction and starting at the lattice site $x$.

What we want is a match of this lattice theory to the continuum theory
such that they produce the same infrared effective theory, up to
small and controlled errors.  The general philosophy of the matching
procedure was developed by Symanzik years ago \cite{Symanzik}.
If the theory were strictly linear then the matching procedure would be
trivial; setting $g_l^2 = g_c^2$, we could reproduce the behavior of the
continuum theory up to $O(a^2)$ errors associated with nonrenormalizable
operators.  However,
the nonlinearity of the theory makes the infrared behavior
depend on the ultraviolet, so if we change the UV modes then it modifies 
the IR effective theory.  Changing $a$ is changing the UV modes, so the
relation between $g^2_c$ and $g^2_l$ will be more complex.  But the
relation between the theories should be analytic  (or at least asymptotic)
in the coupling when it is small.  So the relation can be expanded;
\begin{equation}
g_l^2 = g_c^2 \left( 1 + c_1(a \mu) g_c^2 + c_2(a \mu) g_c^4 + 
	\ldots \right) \, .
\label{g2series}
\end{equation}
To match the theories at $O(a^0)$ we need to know the values of ALL of
the coefficients $c_i$.  To match beyond $O(a^0)$, say at $O(a^2)$, 
we must write down a lattice action with enough parameters to separately
tune the tree level values of all $O(a^2)$ nonrenormalizable operators
consistent with cubic symmetry,
and the appropriate coefficients can each be written as an expansion in
$g_c^2$ as above.  In theories with fermions there are also potential
$O(a)$ nonrenormalizable operator errors.

In the early 1980's it was hoped that the series such as the one in Eq.
(\ref{g2series}) would be rapidly convergent, and perturbative
computation of the first few terms would give a good match between
lattice and continuum theories \cite{Hasenfratz}.  
In fact, the lattice perturbative calculations are very complicated, 
and the convergence of the perturbative series is not very good.  
It is more practical to make nonperturbative
measurements in the lattice theory and find $g_l^2$
by matching their values to continuum (perturbative or measured)
values.  It is also possible to go beyond an $O(a^0)$ matching; for
instance, L{\"u}scher's group has recently been able to match lattice 
Yang-Mills theory with Wilson fermions to the continuum theory, using
an improved lattice action and nonperturbative matching conditions to
remove all $O(a^0)$ and $O(a)$ errors at the nonperturbative level
\cite{Luscher}.  

The idea in 3-D bosonic Yang-Mills Higgs theory is similar, but 
because we are in a lower dimension, the behavior of the expansion is
fundamentally different.  The continuum action, say, when there is a
fundamental complex Higgs field, is
\begin{equation}
\int d^3 x \left( \frac{1}{2} {\rm Tr} F_{ij} F_{ij} 
	+ (D_i \phi)^{\dagger} D_i \phi 
	+ m_{3c}^2 \phi^\dagger \phi + \lambda_{3c} (\phi^\dagger \phi)^2 
	\right) \, .
\end{equation}
The dimensionalities of the fields and couplings are now $[\phi] = {\rm
energy}^{1/2}$, $[A] = {\rm energy}^{1/2}$, $[g^2] = [\lambda] = {\rm
energy}$, and $[m^2_3] = {\rm energy}^2$.  Note that the coupling
constants are dimensionful.  In terms of the 4 dimensional couplings,
$g_3^2 = g_4^2 T$ at tree level.\footnote{Alternately one may maintain 
the 4-D normalization of the fields and couplings, but then a dimensionful
factor, $1/T$, appears in front of the action, so $T$ always appears
along with $g^2$ or $\lambda$ in the loopwise expansion.  This leads to
the same parametric arguments we present here, 
as each $g^2$ appears with a $T$
which has to be balanced by an $a$ for dimensional reasons.}

A lattice version of this theory has as its action
\begin{eqnarray}
a^3 \sum_x &  \Bigg\{ \! \! \! \! \! & 
	\frac{2}{g_{3l}^2 a^4} \sum_{i<j} {\rm Tr} \left( 
	{\bf 1} - P_{ij} \right) + \frac{Z_\phi}{a^2} \sum_i \left(
	(\phi(x) - U_i(x) \phi(x+i))^\dagger (\phi(x) - U_i(x) 
	\phi(x+i)) \right) + \nonumber \\
	& & + m_{3l}^2 Z_\phi \phi^\dagger(x) \phi(x)
	+ \lambda_{3l} Z_\phi^2 
	\left( \phi^\dagger  \phi \right)^2 \Bigg\} \, .
\end{eqnarray}
If the theory were noninteracting then the lattice-continuum relations
would be 
\begin{equation}
g_{3l}^2 = g_{3c}^2 \, , \quad Z_\phi = 1 \, , \quad
	\lambda_{3l} = \lambda_{3c} \, , 
	\quad m_{3l}^2 = m_{3c}^2 \, , 
\end{equation}
and the measured lattice value of the operator insertion
$\langle \phi^\dagger \phi \rangle$
would match the continuum value after lattice zero point fluctuations
were corrected for.  (The only operator insertion we will consider 
in this work is $\langle \phi^\dagger \phi \rangle$.)

The corrections to these relations must again be analytic in $g_3^2$,
$\lambda_3$, and $m_3^2$.  But all of these quantities are dimensionful,
and their dimensionality will be balanced by powers of $a$.  So, for
instance, 
\begin{equation}
g^2_{3l} = g^2_{3c} \left( 1 + c_1 g_{3c}^2 a + c_2 g_{3c}^4 a^2 + 
	c_3 g_{3c}^2 \lambda_c a^2 + \ldots \right) \, .
\end{equation}
The appearance of positive powers of $a$ in this equation is because of
superrenormalizability.  Because all couplings have positive
engineering dimension, the theory can be freed of errors at any
particular order in $a$ by a perturbative caluclation with a finite
number of diagrams.

The only corrections which can occur at $O(a^{-1})$ are a correction
to the mass,
\begin{equation}
\delta m_l^2 \sim g_c^2 a^{-1} \, , \; \lambda_c a^{-1}
\end{equation}
and a constant difference between lattice and continuum values for
$\langle \phi^\dagger \phi \rangle$,
\begin{equation}
\delta \langle \phi^\dagger \phi \rangle_l \sim  a^{-1} \, .
\end{equation}
These are generated by one loop diagrams, and the latter is already
present at zero coupling.  Here and in what follows the $3$ subscript
will be implied, when there can be no confusion.

At $O(a^0)$ or $O(a^0 \ln a \mu)$, the only corrections are of form
\begin{equation}
\delta m_l^2 \sim g_c^4 \, , \; g_c^2 \lambda_c \, , \; \lambda_c^2 \, ,
\end{equation}
and
\begin{equation}
\delta \langle \phi^\dagger \phi \rangle \sim g_c^2 \, , \; \lambda_c \, .
\end{equation}
The coefficients are determined by a two loop calculation and 
may contain $\ln ( \mu a )$.  Knowing all of these corrections allows
one to make a match which is correct in the strict $a \rightarrow 0$
limit.  These counterterms are computed for SU(2) Higgs theory in
\cite{FKRS,Laine} and extended to SU($N$) in \cite{LaineRajantie}.

At $O(a)$, the possible additive corrections 
to $m_l^2$ and $\langle \phi^\dagger \phi \rangle$ are
\begin{equation}
\delta m_l^2 \sim ag_c^6 \, , \; ag_c^4 \lambda_c \, , \; ag_c^2
	\lambda_c^2 \, , \; a \lambda_c^3 \, ,
\end{equation}
and
\begin{equation}
\delta \langle \phi^\dagger \phi \rangle 
	\sim ag_c^4 \, , \; ag_c^2 \lambda_c \, , \;  a \lambda_c^2 \, .
\end{equation}
These arise from 3 loop diagrams.  The other possible corrections are
\begin{eqnarray}
m_l^2 - m_c^2 & \sim & a g_c^2 m_c^2 \, \; a \lambda_c m_c^2 \, , \\
\langle \phi^\dagger \phi \rangle_l - \langle \phi^\dagger \phi 
	\rangle_c & \sim &  a g_c^2 
	\langle \phi^\dagger \phi \rangle_c \, , \;
	a \lambda_c \langle \phi^\dagger 
	\phi \rangle_c \, , \; a m^2_c \, , \\
Z_\phi - 1 & \sim & a g_c^2 \, , \; a \lambda_c \, , \\
\delta g_l^2 & \sim & a g_c^4 \, , \\
\delta \lambda_l & \sim & a g_c^4 \, , \; a g_c^2 \lambda_c \, , \;
	a \lambda_c^2 \, .
\end{eqnarray}
These corrections are potentially generated by one loop diagrams.

We should also check whether nonrenormalizable operators are induced;
the terms we might possibly need to include 
in the tree action are of form
\begin{equation}
(a g_c^2 \, , \; a \lambda_c ) (\phi^\dagger \phi)^3 \, , \;
a F^2 \phi^\dagger \phi \, , \; 
a \phi^\dagger \phi (D_i \phi)^2 \, , \;
a (\phi^\dagger \phi)^8 \, .
\end{equation}
But any diagram which generates one of these contains too many powers of
$g^2$ or $\lambda$, which must be balanced by more powers of $a$.  Even
at one loop, none of these terms are generated below $O(a^3)$.  So the
corrections listed are exhaustive at $O(a)$.

It would be nice to compute all $O(a)$ corrections.  But we see that two
of these,  additive corrections to $m_l^2$ and 
$\langle \phi^\dagger \phi \rangle$
respectively, involve 3 loop computations, which are prohibitively
difficult.  Hence a complete analytic treatment of all $O(a)$
corrections appears unrealistic.  Nevertheless, we accomplish something
worthwhile by computing the remaining coefficients.  This is because
they influence and report the strength of the phase transition, while 
the additive corrections to $m_l^2$ and $\langle \phi^\dagger \phi \rangle$
do not.  Computing the others eliminates $O(a)$ errors in jumps in order
parameters, the surface tension, the latent heat, the profile of the
phase interface, the temperature range of metastability, etc.  Only the
$\overline{\rm MS}$ value of $m_{3c}^2$ at $T_c$ and the absolute 
$\overline{\rm MS}$ value of $\langle \phi^\dagger \phi \rangle$
in one phase are determined with less accuracy.  Further, it may be much
easier to determine the (uncalculated) coefficients for the 3-loop
$m_l^2$ and $\langle \phi^\dagger \phi \rangle$
corrections by Monte-Carlo methods if all other $O(a)$ errors are
eliminated, and it may be easier to compute these than it would be to
eliminate all $O(a)$ errors in all quantities by extrapolation.

Finally we should point out that the 3-D values of $m_3^2$ and 
$\langle \phi^\dagger \phi \rangle$
are known with less accuracy in terms of the 4-D parameters (and
hence physical measurables) as $g_3^2$, $\lambda_3$, and wave
function normalizations.  Writing
\begin{equation}
\label{3d4dforg}
g_{3c}^2 = g_4^2(\mu) T \left( 1 + c_1(\mu/T) g_4^2 + c_2(\mu/T) g_4^4 
	+ \ldots \right) \, ,
\end{equation}
a 1 loop calculation of the dimensional reduction step already
establishes the value of $c_1$, and the relative uncertainty in $g_3^2$
is $O(\alpha^2)$.  On the other hand, a 1 loop determination of
$m_{3c}^2 / g_{3c}^4$ leaves an $O(1)$ error, and the 2 loop
determination, which is presently state of the art, leaves an
$O(\alpha)$ error.\footnote{This is for the integration over the
``superheavy'' scale, ie nonzero Matsubara frequencies.  
If one also integrates out zero Matsubara frequency modes which take on
$O(gT)$ masses, the ``heavy'' scale, then $\alpha$ in this argument
becomes $\alpha^{1/2}$.}
Improving this by doing the 3 loop computation would
be of comparable difficulty to the 3 loop calculation of the lattice
continuum match.  If we make the parametric estimate that $g_{3c}^2 a
\sim \alpha$, ie $a \sim 1/T$, then the $O(a)$ calculation of $\delta
m_{3l}^2$ is unjustified without the 3 loop dimensional reduction
calculation.  But the $O(a)$ calculation of the other quantities is
parametrically justified.

It is possible in principle to continue the lattice to continuum match
to arbitrary order in $a$.  At any finite order, the lattice action
requires a finite number of terms and the matching requires the
calculation of a finite number of diagrams.  In practice the complexity
rises very fast beyond $O(a)$.  To remove all $O(a^2)$ errors in
measurables describing the strength of the phase transition, one must
expand the lattice action to contain terms which can be tuned to remove
all tree level $O(a^2)$ nonrenormalizable operators.  This changes the
Feynman rules of the perturbation theory.  One will also need to compute
corrections to the wave functions and couplings at two loops, which
involves over 200 topologically distinct
diagrams.  We will content ourselves with $O(a)$ improvement here.

\section{Improvement in SU($N$) theory}
\label{details}

Our goal is, given a continuum SU($N$) plus real adjoint Higgs plus
complex fundamental Higgs theory in 3 dimensions, to write down the
lattice theory which is equivalent up to $O(a)$ additive errors in
masses and expectation values of operator insertions and $O(a^2)$ errors
in wave functions, couplings, multiplicative corrections to masses and
operator insertions, and nonrenormalizable operators.

We denote the adjoint Higgs field as $\Phi = \Phi^a T^a$ and the
fundamental Higgs field as $\phi$.  Our group conventions are standard,
ie ${\rm Tr} T^a T^b = (1/2) \delta_{ab}$.  It is to be understood that
$D_i$ acting on $\Phi$ is the adjoint covariant derivative and $D_i$
acting on $\phi$ is the fundamental covariant derivative.  
The Lagrangian density of the continuum theory we consider is
\begin{eqnarray}
{\cal L}_{\rm cont}& = & 
	\frac{1}{2} {\rm Tr} F_{ij} F_{ij} + {\rm Tr} ( D_i \Phi D_i \Phi) 
	+ (D_i \phi)^{\dagger} (D_i \phi) + \nonumber \\
&& + m_\Phi^2 {\rm Tr} \Phi^2 + m_\phi^2 \phi^\dagger \phi + 
	\lambda_1 ( {\rm Tr} \Phi^2 )^2 + \lambda_2 {\rm Tr} \Phi^4
	+ \lambda ( \phi^\dagger \phi)^2 + \nonumber \\
&&	+ h_1 {\rm Tr} \Phi^2 \phi^\dagger \phi + 
	+ \frac{h_2}{2} d_{abc} \Phi^a \Phi^b \phi^\dagger T^c \phi \, .
\end{eqnarray}
Note that there are two possible adjoint field self-interactions and two
possible interactions between fundamental and adjoint Higgs fields.
In SU(2) and in SU(3), ${\rm Tr} \Phi^4 = (1/2) ({\rm Tr} \Phi^2)^2$.
In SU(2) this is because $d_{abc} = 0$, and in SU(3) it is because of
the relation, special to SU(3), that
\begin{equation}
d_{abe} d_{cde} + d_{ace} d_{bde} + d_{ade} d_{bce} = \frac{1}{3}
	\left( \delta_{ab} \delta_{cd} + \delta_{ac} \delta _{bd} + 
	\delta_{ad} \delta_{bc} \right) \; {\rm for \; SU(3)} \, .
\end{equation}
Hence, in these theories the two adjoint Higgs field self-interactions
are dependent, and only the combination $(\lambda_1 + \lambda_2 / 2)$ is
important.  Also note that the $h_2$ term can be dropped for SU(2),
because $d_{abc} = 0$ in that case.

A choice for the Lagrangian density for the lattice theory 
which is general enough to allow $O(a)$ corrections is
\begin{eqnarray}
{\cal L}_{\rm latt}(x) & = & \frac{2}{Z_g g^2 a^4} \sum_{i<j} {\rm Tr}
	( {\bf 1} - P_{ij}(x) ) 
	+ \frac{2 Z_\phi}{a^2} \sum_i [ \phi^\dagger(x) \phi(x) - 
	\phi^\dagger(x) U_i(x) \phi(x+i) ]  + \nonumber \\
	& & + \frac{2Z_\Phi}{a^2} \sum_i
	[ {\rm Tr} \Phi^2(x) - {\rm Tr} \Phi(x) U_i(x) \Phi(x+i)
	U^\dagger_i(x) ] + \nonumber \\
	& & + (\lambda + \delta \lambda) Z_\phi^2 (\phi^\dagger \phi)^2
	+ (\lambda_1 + \delta \lambda_1) Z_\Phi^2 ( {\rm Tr} \Phi^2 )^2
	+ (\lambda_2 + \delta \lambda_2) Z_\Phi^2 {\rm Tr} \Phi^4 + 
	\nonumber \\
	& & + ( h_1 + \delta h_1 ) Z_\phi Z_\Phi \phi^\dagger \phi 
	{\rm Tr} \Phi^2 + \frac{(h_2 + \delta h_2 )}{2} Z_\phi Z_\Phi 
	d_{abc} \Phi^a \Phi^b \phi^\dagger T^c \phi + \nonumber \\ 
	& & + \delta m_\phi^2 Z_\phi \phi^\dagger \phi + 
	\delta m_\Phi^2 Z_\Phi {\rm Tr} \Phi^2 +  \nonumber \\
	& & + \left[ m_\phi^2  \quad m_\Phi^2 \right] Z_m 
	\left[ Z_\phi \phi^\dagger \phi \quad Z_\Phi {\rm Tr}
	\Phi^2 \right]^T \, .
\label{lattlagrangian}
\end{eqnarray} 
The couplings and masses here are the continuum theory values, and the
coefficients $Z_g$, $Z_\Phi$, $Z_\phi$, $\delta \lambda_i$, 
$\delta m^2_i$, and $Z_m$ perform the $O(a)$ corrections.  $Z_m$ is a $2
\times 2$ matrix, which differs at $O(a)$ from the identity matrix.

Further, to convert measured values of operator insertions to the
equivalent continuum values, we need an addative and a multiplicative
renormalization of the operator insertions,
\begin{equation}
\left[ \begin{array}{c} \langle \phi^\dagger \phi \rangle \\ 
	\langle {\rm Tr} \Phi^2 \rangle \\
	\end{array} \right]_{\rm continuum } = \left[ \begin{array}{c}
	\delta \langle \phi^\dagger \phi \rangle \\
	\delta \langle {\rm Tr} \Phi^2 \rangle \\ \end{array} \right]
	+ Z_{\rm OP} \left[ \begin{array}{c} 
	\langle \phi^\dagger \phi \rangle \\ 
	\langle {\rm Tr} \Phi^2 \rangle \\
	\end{array} \right]_{\rm latt, \; measured} \, .
\end{equation}
Here $Z_{\rm OP}$ is also a matrix, with $O(a)$ off diagonal elements
which account for mixing between the operator insertions.
All notations are chosen to follow \cite{LaineRajantie}; to convert to the
notation of \cite{Oapaper}, change the signs on the counterterms in the
scalar effective potential, introduce $\lambda_{L,i} = 4 \lambda_i /
g^2$, rescale the scalar fields by a factor of $g^2 a T/4$ and the
$m^2$ by a factor of $a^2$, and write $ 4 / g^2 a \equiv \beta_L$.
Also, $Z_g$ was called $Z_A^{-1}$ there.

Neither $Z_\phi$, $Z_\Phi$, nor $Z_{\rm OP}$ are separately gauge
invariant or physical; one may for instance rescale 
the $\Phi$ field by a factor of $\eta = 1 + O(a)$ and change $Z_\Phi$ by
$\eta^{-2}$ and the right column of $Z_{\rm OP}$ by 
$\eta^{-2}$ without changing anything.  It is the matrix
$Z_{\rm OP} \times {\rm Diag} [ Z_\phi^{-1} \; \; Z_\Phi^{-1}]$ 
which is gauge invariant.  This combination equals $Z_m$, a statement
which will remain true at all orders.

Similarly, while $Z_g$ is gauge
invariant, if one tries to relate the 3-D theory gauge field $A$ to the
lattice gauge field $A$ defined by $U_i(x) = \exp( -i a g T^a
A^a_i(x+i/2))$, one will find the relation contains a non-gauge 
invariant $O(a)$ correction, for the same reason that the 4-D continuum
gauge field wave function renormalization is not gauge invariant, while
the renormalization of the coupling constant is.  This is only important
if one is interested in measuring $A$ field correlators in some gauge,
and we will not compute the $A$ field wave function correction here.

\begin{figure}[t]
\centerline{\mbox{\psfig{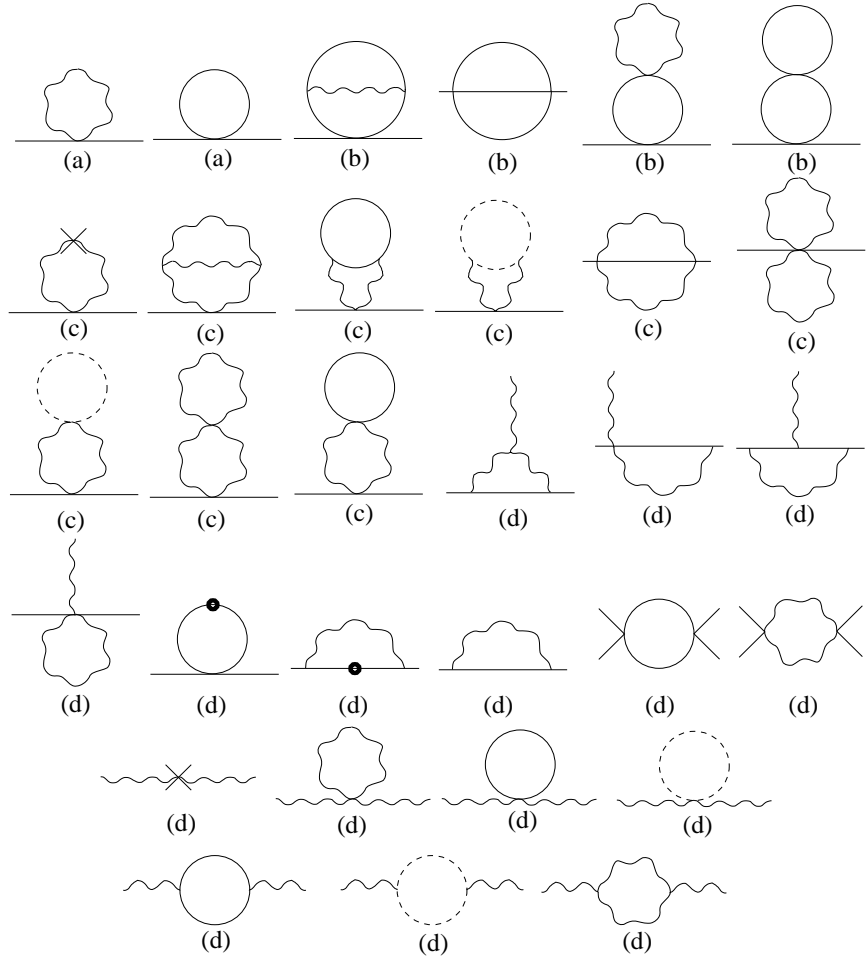}}}
\caption{\label{alldiagrams} The topologically distinct diagrams 
needed for the Landau gauge calculation.  Wavy lines are gauge 
propagators, solid lines are scalars, and dotted lines are ghosts.  
A cross is a self energy correction from the measure and
a blot is a $\phi^2$ operator insertion.}
\end{figure}

To determine all of the required corrections, one must calculate all
1 loop self-energy insertions for all fields to $O(p^2)$, 
all 1 loop $O(p)$ corrections to one scalar-gauge vertex, all 1 loop
scalar 4-point corrections at zero external momentum, all 1 loop
corrections to the insertion of a $\phi^\dagger \phi$ or ${\rm Tr}
\Phi^2$ operator on a zero momentum scalar line, all 1 and 2 loop scalar
self-energy insertions at zero momentum, and all 1 and 2 loop vacuum
diagrams with $\phi^\dagger \phi$ or ${\rm Tr \Phi^2}$ insertions.  The
topologically distinct diagrams required in Landau gauge
are presented in Figure
\ref{alldiagrams}.  In a general gauge, additional diagrams are needed.
The Feynman rules are spread between \cite{Rothe,Laine,Oapaper}.

All of the diagrams have been computed previously: $(a)$ at 
$O(p^0)$ in \cite{KRSearly}, $(b)$ in \cite{FKRS}, $(c)$ in
\cite{Laine},\footnote{The computation there is not organized in terms
of two loop scalar self-energy insertions, but it is equivalent.} 
and $(a)$ at $O(p^2)$, $(d)$ in \cite{Oapaper}.  Our
task here is purely combinatoric.  We must include 2 types of scalars
and compute the appropriate SU($N$) group factors.  All the group
theoretic identities which are needed appear in Appendix A 
of \cite{Rajantie}.  The converse is also true.

Rather than present the full combinatorial details of the calculation,
we will skip to the results.  Two numerical constants
appear, and their values are $\Sigma = 3.175911536$, $\xi =
0.152859325$.  Also, when comparing the dimensionality of the right and
lefthand sides, remember that the coupling constants $g^2$, $\lambda_i$
are in 3-D notation and correspond to $T$ times the dimensionless 4-D
values (at some renormalization point).  The coupling renormalizations 
are\footnote{We choose to report $Z_g^{-1} - 1$ because
at two loops this quantity is only changed by two loop self-energy
diagrams and vertex corrections, ie it absorbs interated insertions of 1
loop self-energy diagrams on propagators.  But at $O(a)$ accuracy we
could equally have written Eq. (\protect{\ref{gaugerenorm}}) as a
correction to $1 - Z_g$.}
\begin{eqnarray}
\label{gaugerenorm}
Z_g^{-1} - 1 & = & \frac{g^2 a}{4 \pi} 
	\left( \frac{2 \pi}{9N} (2N^2 - 3) + 
	N \left( \frac{37 \xi}{12} - \frac{\pi}{9} \right) +
	N \left[ \frac{\Sigma}{24} - \frac{\xi}{6} \right] + 
	\left\{ \frac{\Sigma}{24} - \frac{\xi}{6} \right\}
	\right) \, , \\
\delta \lambda & = & \frac{a}{4\pi} \Bigg( \frac{N^3 + N^2 - 4N + 2}
	{4N^2} g^4 \xi + (2N + 8) \lambda^2 \xi + \frac{N^2 - 1}{4} 
	h_1^2 \xi + \nonumber \\
	& & \qquad + \frac{(N^2 - 4)(N-1)}{8 N^2} h_2^2 \xi - 
	\frac{N^2 -1}{6N} (18 \xi + \Sigma) g^2 \lambda \Bigg) \, , \\
\delta \lambda_1 & = & \frac{a}{4\pi} \Bigg( 3 g^4 \xi + (N^2 + 7) 
	\lambda_1^2 \xi + \frac{4N^2 - 6}{N} \lambda_1 \lambda_2 \xi +
	\nonumber \\ & & \qquad + \frac{3N^2 + 9}{N^2} \lambda_2^2 \xi 
	+ \frac{N}{2} h_1^2 \xi - \frac{1}{2N} h_2^2 \xi - 
	\frac{N}{3} (18 \xi + \Sigma) g^2 \lambda_1 \Bigg) \, , \\
\delta \lambda_2 & = & \frac{a}{4\pi} \Bigg( N g^4 \xi 
	+ 2 \frac{N^2 - 9}{N} \lambda_2^2 \xi 
	+ 12 \lambda_1 \lambda_2 \xi + \frac{1}{2} h_2^2 \xi
	- \frac{N}{3} (18 \xi + \Sigma) g^2 \lambda_2 \Bigg) \, , \\
\delta h_1 & = & \frac{a}{4 \pi} \Bigg( 2 g^4 \xi 
	+ (N^2 + 1) h_1 \lambda_1 \xi + (2N + 2) h_1 \lambda \xi 
	+ 2 h_1^2 \xi + \frac{N^2 - 4}{N} h_2^2 \xi + \nonumber \\
	& & \qquad + \frac{2N^2 - 3}{N} \lambda_2 h_1 \xi
	- \frac{3N^2 - 1}{12 N} (18 \xi + \Sigma) 
	g^2 h_1 \Bigg) \, , \\
\delta h_2 & = & \frac{a}{4 \pi} \Bigg( N g^4 \xi + h_2 \left(
	2 \lambda + 2 \lambda_1 + \frac{N^2 - 6}{N} \lambda_2
	+ 4 h_1 + \frac{N^2 - 12}{2N} h_2 \right) \xi 
	- \nonumber \\ & & \qquad - \frac{3N^2 - 1}{12 N} 
	(18 \xi + \Sigma) g^2 h_2 \Bigg) \, .
\end{eqnarray}
For the theory without the adjoint Higgs field, drop the term with
square brackets in Eq. (\ref{gaugerenorm}), set $\lambda_1 = \lambda_2 =
h_1 = h_2 = 0$ in the equation for $\delta \lambda$, and ignore the
equations for $\delta \lambda_1$ etc.  
Putting $N=2$, we recover the results
of \cite{Oapaper}, though the notation there is somewhat different.
For the theory without the fundamental
Higgs, drop the term in curly brackets in Eq. (\ref{gaugerenorm}), set
$\lambda = h_1 = h_2 = 0$, and ignore the equations for their
corrections.  

The fundamental scalar contribution to Eq. (\ref{gaugerenorm}) is wrong
in \cite{Oapaper} by a factor of $1/2$, both for SU(2) and U(1).  
This error is not numerically
important; even in SU(2) it leads to a $3\%$ error 
in $Z_g^{-1}-1$.  Appendix C of that paper, which
treats the case with both Higgs fields for $N=2$, has
several typographical omissions of factors of 
$\Sigma/4\pi$ and $\xi/4\pi$, and also the $g^4$
contributions to $( \delta \lambda_1 + \delta \lambda_2/2 )$ 
and to $\delta h_1$ are both off by
a factor of 2 there, due to a combinatorial error.  These are the only
combinatorial mistakes in that paper.\footnote{However, early preprint
versions had further mistakes which were corrected before publication.}

Before continuing, we mention two checks on the calculation so far.
First, both for $N=2$ and $N=3$,  the combination $(\lambda_1 +
\lambda_2 / 2)$ is important, 
not $\lambda_1$ and $\lambda_2$ separately.
Checking, one finds that $(\delta \lambda_1 + \delta \lambda_2 / 2)$
is fixed when one varies $\lambda_1$ keeping $(\lambda_1 + 
\lambda_2 / 2)$ fixed, both for $N=2$ or $N=3$.
Also, $h_2$ should have no influence when $N=2$.  Indeed, its
contribution to the rescaling of other couplings vanishes for $N=2$.

Now let us present $Z_\Phi$, $Z_\phi$, and $Z_{\rm OP}$.  Since they are
not separately gauge invariant, we will present them in Landau gauge.
In this gauge, at one loop, the corrections to wave functions all go as
$g^2$ and the corrections to $Z_{\rm OP}$ depend on scalar
self-couplings.  This statement does not persist at 2 loops.  The
results are
\begin{eqnarray}
Z_\phi - 1 & = & \frac{g^2 a}{4 \pi} \frac{N^2 - 1}{12 N} 
	(18 \xi + \Sigma) \, , \\
Z_\Phi - 1 & = & \frac{g^2 a}{4 \pi} \frac{N}{6} (18 \xi + \Sigma)
	\, , \\
Z_{\rm OP} - {\bf 1} & = & \frac{\xi a}{4 \pi} \left[
	\begin{array}{cc} (2N+2) \lambda & N h_1 \\
	\frac{N^2 - 1}{2} h_1 & (N^2 + 1) \lambda_1 
	+ \frac{2N^2 - 3}{N} \lambda_2 \\ \end{array} \right] \, .
\end{eqnarray}
Here ${\bf 1}$ is the $2 \times 2$ identity matrix.
Varying the gauge changes the wave function corrections and the 
on diagonal elements of $Z_{\rm OP}$, but not $Z_m$.
The off diagonal elements of $Z_{OP}$ are also gauge
invariant at $O(a)$, but this will not persist at higher loop orders.

It remains to present $\delta m^2_\phi$, $\delta m^2_\Phi$, 
$\delta \langle \phi^\dagger \phi \rangle$, 
and $\delta \langle {\rm Tr} \Phi^2
\rangle$.  The one loop corrections are \cite{LaineRajantie}
\begin{eqnarray}
\label{dm1_1loop}
\delta m_\phi^2 ({\rm one \; loop}) & = & 
	- \frac{\Sigma}{4 \pi a} \left( \frac{N^2 - 1}{N} g^2 + 2(N+1)
	\lambda + \frac{N^2 - 1}{2} h_1 \right) \, , \\
\label{dm2_1loop}
\delta m_\Phi^2 ({\rm one \; loop}) & = & 
	- \frac{\Sigma}{4 \pi a} \left( 2 N g^2 + N h_1
	+ (N^2 + 1) \lambda_1 + \frac{2N^2 - 3}{N} \lambda_2 
	\right) \, , \\
\delta \langle \phi^\dagger \phi \rangle ({\rm one \; loop}) & = & - N 
	\left( \frac{\Sigma}{4 \pi a} + \frac{ \xi m_\phi^2 a}{4 \pi}
	\right) \, , \label{dphisq1loop}\\
\delta \langle {\rm Tr} \Phi^2 \rangle ({\rm one \; loop}) 
	& = & - \frac{N^2 - 1}{2} 
	\left( \frac{\Sigma}{4 \pi a} + \frac{ \xi m_\Phi^2 a}{4 \pi}
	\right) \, . \label{dPhisq1loop}
\end{eqnarray}
We have included the $O(a)$, $m^2$ dependent correction to the operator
insertion expectation values.  For the case where there is one Higgs
field but not the other, set $h_1=0$.  Also note that our
$h_1$ differs by a factor of 2 in normalization
from the parameter $h_3$ appearing in \cite{Laine}.  

The two loop corrections have been computed in \cite{LaineRajantie}, for
the case where there is an adjoint scalar, or a fundamental scalar, but
not both.  For our case, we must add new
terms when there are two types of scalars, and contributions from
inserting $O(a)$ counterterms into 1 loop diagrams.

The two loop contributions listed in \cite{LaineRajantie} are
\begin{eqnarray}
\delta m_\phi^2 & += & - \frac{N^2 - 1}{16 \pi^2}
	\Bigg\{ \left[ g^4 \frac{4N^2 - N + 3}{4N^2} + 2 \lambda g^2 
	\frac{N+1}{N} - 4 \lambda^2 \frac{1}{N-1} \right] \left(
	\ln \frac{6}{a \mu} + \zeta \right) + \nonumber \\
	& & \quad + 2 \lambda g^2 
	\frac{N+1}{N}\left( \frac{\Sigma^2}{4} -
	\delta \right) + g^4 \frac{1}{4N^2} \Bigg[ \frac{4N^2 - 1}{4} 
	\Sigma^2 + \frac{3N^2 - 8}{3} \pi \Sigma + N^2 + \nonumber \\
	& & \quad + 1 - 4 N(N+1) \rho - 2(3N^2-1) \delta + 2 N^2
	(2 \kappa_1 - \kappa_4 ) \Bigg] \Bigg\} \, , \\
\delta m_\Phi^2 & += & - \frac{1}{16 \pi^2}
	\Bigg\{ \Bigg[ 2Ng^2 \left( (N^2+1) \lambda_1 
	+ \frac{2N^2-3}{N} \lambda_2 \right) - 2 (N^2+1) \lambda_1^2
	 - \nonumber \\
	& & \qquad - \frac{4(2N^2-3)}{N} \lambda_1 \lambda_2
	- \frac{N^4 -6N^2 +18}{N^2} \lambda_2^2 \Bigg]
	\left( \ln \frac{6}{a\mu} + \zeta \right) + \nonumber \\
	& & \qquad + 2Ng^2 \left(
	\frac{\Sigma^2}{4} - \delta \right) \times 
	 \left( (N^2+1) \lambda_1 + \frac{2N^2-3}{N} \lambda_2
	\right) + \nonumber \\
	& & \qquad + g^4 N^2 \left[ \frac{5 \Sigma^2}{8} + 
	\frac{3N^2-8}{6N^2} \pi \Sigma - 4 (\delta + \rho) + 2 
	\kappa_1 - \kappa_4 \right] \Bigg\} \, , \\
\delta \langle \phi^\dagger \phi \rangle & += & - \frac{g^2}{16 \pi^2} 
	(N^2 - 1) \left( \ln \frac{6}{a \mu} + \zeta + 
	\frac{\Sigma^2}{4} - \delta \right) \, , \\
\delta \langle {\rm Tr} \Phi^2 \rangle & += & -  \frac{g^2}{16 \pi^2} 
	N (N^2 - 1) \left( \ln \frac{6}{a \mu} + \zeta + 
	\frac{\Sigma^2}{4} - \delta \right) \, ,
\end{eqnarray}
where the newly introduced constants are $\zeta = .08849,$ 
$\delta = 1.942130$, $\rho = -.313964$, $\kappa_1 = .958382$, $\kappa_4 =
1.204295$.  Here and in the following we will write $+=$ to show that
these contributions are to be added to those listed previously.

In the case where both Higgs fields are present, there are added
contributions to the mass counterterms of
\begin{eqnarray}
\delta m_\phi^2 & += & - \frac{N^2-1}{16 \pi^2} \Bigg\{
	\left( - \frac{1}{2} h_1^2 - \frac{N^2-4}{4N^2} h_2^2
	+ N h_1 g^2 - \frac{1}{4} g^4 \right) \times \left(
	\ln \frac{6}{a \mu} + \zeta \right) + \nonumber \\
	& & \qquad \qquad 
	+ \left( \frac{\Sigma^2}{4} - \delta \right) N
	g^2 h_1 - \rho g^4 \Bigg\} \, , \\
\delta m_\Phi^2 & += & - \frac{1}{16 \pi^2} \Bigg\{
	\left( - N h_1^2 - \frac{N^2-4}{2N} h_2^2 + (N^2-1) h_1 g^2
	- \frac{N}{2} g^4 \right) \times \left(
	\ln \frac{6}{a \mu} + \zeta \right) + \nonumber \\
	& & \qquad \qquad
	+ \left( \frac{\Sigma^2}{4} - \delta \right) 
	(N^2-1) h_1 g^2 - 2N \rho g^4 \Bigg\} \, .
\end{eqnarray}
At $N=2$ these match the results of \cite{Laine}.  Do not include these
in the theory with only a fundamental or an adjoint Higgs, but not both.

In addition, there are contributions at $O(a^0)$ 
arising from $O(a)$ conterterm insertions into diagrams which give
$O(1/a)$ divergences.  These lead to the following corrections:
\begin{eqnarray}
\delta m_\phi^2 & += & - \frac{\Sigma}{4 \pi a} \left( 
	\frac{N^2-1}{N} g^2 (Z_g - 1) + 2 (N+1) \delta \lambda + 
	\frac{N^2-1}{2} \delta h_1 \right) \, , \label{newmcounter1}\\
\delta m_\Phi^2 & += & - \frac{\Sigma}{4 \pi a} \left( 
	2N g^2 (Z_g - 1) + N \delta h_1 + (N^2+1) \delta \lambda_1 + 
	\frac{2N^2-3}{N} \delta \lambda_2 \right) \, , 
	\label{newmcounter2}\\
\left[ \begin{array}{c} \delta \langle \phi^\dagger \phi \rangle \\
	\delta \langle {\rm Tr} \Phi^2 \rangle \\ \end{array} \right]
	& += & \Big( Z_m
	- {\bf 1} \Big) \left[ \begin{array}{c} 
	\delta \langle \phi^\dagger \phi \rangle \\
	\delta \langle {\rm Tr} \Phi^2 \rangle \\ \end{array} \right]
	 ({\rm 1 \; loop}) \, .
	\label{phicounter}
\end{eqnarray}
The $O(a)$ counterterms appearing here were given earlier, and as usual
the term in the parenthesis in Eq. (\ref{phicounter}) is to be
interpreted as a $2 \times 2$ matrix.
It should be clear how to pare these equations down to the case where
there is only one type of Higgs field.

This completes the calculation of all counterterms needed for the $O(a)$
improvement of the theory, except for the additive $O(a)$ corrections to
the masses and operator insertions which arise at 3 loops, which we will
not calculate, as advertized.

What should you do, though, if you have already taken data without
including these corrections in the Lagrangian?  That is, what if you
have data for a theory which naively has a lattice spacing $a_n$, no
wave function or $g$ renormalization, and naive couplings $\lambda_n$,
etc?  Then you should figure out what theory you were ``really'' looking
at, and reinterpret the results accordingly.
By demanding that the ``naive'' action
\begin{eqnarray}
a_n^3 \sum_x & \Bigg\{ \! \! \! \! \! \! & 
	\frac{2}{g^2 a_n^4} \sum_{i<j} {\rm Tr}
	( {\bf 1} - P_{ij}(x) ) 
	+ \frac{2}{a_n^2} \sum_i [ \phi_n^\dagger(x) \phi_n(x) - 
	\phi_n^\dagger(x) U_i(x) \phi_n(x+i) ]  + \nonumber \\
	& & + \frac{2}{a_n^2} \sum_i
	[ {\rm Tr} \Phi_n^2(x) - {\rm Tr} \Phi_n(x) U_i(x) \Phi_n(x+i)
	U^\dagger_i(x) ] + \nonumber \\
	& & + \lambda_n (\phi_n^\dagger \phi_n)^2
	+ \lambda_{1n} ( {\rm Tr} \Phi_n^2 )^2
	+ \lambda_{2n} {\rm Tr} \Phi_n^4 + \nonumber \\
	& & + h_{1n} \phi_n^\dagger \phi_n 
	{\rm Tr} \Phi_n^2 + \frac{h_{2n}}{2} d_{abc} 
	\Phi_n^a \Phi_n^b \phi_n^\dagger T^c \phi_n + \nonumber \\ 
	& & + (m_{\phi,n}^2 + \delta m_{\phi,n}^2 ) 
	\phi_n^\dagger \phi_n + \nonumber \\ 
	& & + ( m_{\Phi,n}^2 + \delta m_{\Phi,n}^2 ) 
	{\rm Tr} \Phi_n^2 \Bigg\}
\end{eqnarray} 
be equal to $a^3 \sum_x {\cal L}_{\rm latt}(x)$ of Eq.
(\ref{lattlagrangian}), one finds that the naive values labeled with the
$n$ subscript are related to the $O(a)$ corrected values by
\begin{eqnarray}
a & = & Z_g^{-1} a_n \, , \\
\phi & = & ( Z_g Z_\phi^{-1} )^{1/2} \phi_n \, \\
\Phi & = & ( Z_g Z_\Phi^{-1} )^{1/2} \Phi_n \, \\
\lambda & = & Z_g \lambda_n - \delta \lambda \, , \\
\lambda_1 & = & Z_g \lambda_{1n} - \delta \lambda_1 \, , \\
\lambda_2 & = & Z_g \lambda_{2n} - \delta \lambda_2 \, , \\
h_1 & = & Z_g h_{1n} - \delta h_1 \, , \\
h_2 & = & Z_g h_{2n} - \delta h_2 \, , \\
\delta m_\phi^2 & = & Z_g^2 \delta m_{\phi,n}^2 \, , \\
\delta m_\Phi^2 & = & Z_g^2 \delta m_{\Phi,n}^2 \, , \\
\left[ \begin{array}{c} m^2_\phi \\ m^2_\Phi \\ \end{array} \right]
	& = & Z_g^2 Z_m^{-1} \left[ \begin{array}{c}
	m^2_{\phi,n} \\ m^2_{\Phi,n} \\ \end{array} \right] \, .
\end{eqnarray}
Note that, up to $O(a)$ corrections, $\delta m_\phi^2$ computed using 
$a_n$ and the naive scalar self-couplings, and without including 
Eq. (\ref{newmcounter1}), differs from  $\delta m_\phi^2$ computed 
using the improved values and including Eq. (\ref{newmcounter1}), 
by precisely a factor of $Z_g^{-2}$, just what is required 
above.\footnote{To see this, note that one factor of $Z_g$ comes from
using $a$ rather than $a_n$ in Eqs. 
(\protect{\ref{dm1_1loop}},\protect{\ref{dm2_1loop}}) 
and the other comes because Eqs.
(\protect{\ref{newmcounter1}},\protect{\ref{newmcounter2}}) 
multiply the gauge part by $Z_g$, while the scalar part goes as $\lambda
+\delta \lambda = Z_g \lambda_n$.}
This just says that the mass renormalization computed in
\cite{Laine,LaineRajantie} is the right one if you use the ``naive''
couplings and wave functions.  

Next we should relate the $O(a)$ corrected value of the continuum
operator insertion expectation values, through the $O(a)$ corrected
lattice values, to the uncorrected lattice values.  We have
\begin{eqnarray}
\left[ \begin{array}{c} \langle \phi^\dagger \phi \rangle \\ 
	\langle {\rm Tr} \Phi^2 \rangle \\
	\end{array} \right]_{\rm continuum } & = & 
	\left[ \begin{array}{c}
	\delta \langle \phi^\dagger \phi \rangle \\
	\delta \langle {\rm Tr} \Phi^2 \rangle \\ \end{array} \right]
	+ Z_{\rm OP} \left[ \begin{array}{c} 
	\langle \phi^\dagger \phi \rangle \\ 
	\langle {\rm Tr} \Phi^2 \rangle \\
	\end{array} \right]_{\rm latt, \; measured} \nonumber \\
	& = & Z_g Z_{m} \left[ \begin{array}{cc} 
	\langle \phi^\dagger_n \phi_n \rangle_{\rm meas} 
	+ ( \delta \langle
	\phi^\dagger \phi \rangle)_n  \\
	\langle {\rm Tr} \Phi^2_n \rangle_{\rm meas} 
	+ ( \delta \langle {\rm Tr} \Phi^2 \rangle)_n \\
	\end{array} \right] \, .
\end{eqnarray}
Again, using $a_n$ in Eqs. (\ref{dphisq1loop},\ref{dPhisq1loop}) 
and skipping Eq. (\ref{phicounter}) leads to a
value for the naive counterterm which matches the corrected counterterm
in the above.  So if one has ``naive'' measured values for condensates,
with the additive counterterms from \cite{FKRS,Laine,LaineRajantie}
already taken out, then one applies the $O(a)$ corrections by
multiplying by $Z_g Z_m$, where $Z_m$ is to be understood
as a $2 \times 2$ matrix, as usual.

So to apply $O(a)$ corrections to data taken using a ``naive'' lattice
Lagrangian, interpret the true lattice spacing as $Z_g^{-1}a_n$, the
true scalar self-couplings as $Z_g \lambda_n - \delta \lambda$, the true
mass as $m^2 = Z_m^{-1} Z_g^2 m_n^2$, and the true value of order
parameters as $Z_m Z_g$ times the naive values.  Note that $Z_g < 1$
and $(Z_g-1) \lambda - \delta \lambda < 0$, so jumps in order parameters
are generally smaller than the ``naive'' values and the true scalar 
self-couplings are generally smaller than the ``naive'' values.  Using
the ``naive'' values tends to over-report the strength of the phase
transition.  Also remember that the mass correction
should only be used on the difference in 3-D mass
parameter between two trials with the same lattice spacing and scalar
couplings, and the order parameter correction should only be used on 
the difference in the order parameter between phases
or between trials with different 3-D masses (and here one should also
be careful to include the mass dependent term in Eqs.
(\ref{dphisq1loop},\ref{dPhisq1loop})).  This is because we have not
computed the $O(a)$ additive corrections to the mass and order parameter
counterterms. 

To correct the latent heat and the surface tension, note that they are
energies per unit volume and area, respectively; the $O(a)$
corrections come about from the relation between $a$ and $a_n$.
Similarly, correlation lengths are corrected by converting from lattice
units to physical units using $a$ rather than $a_n$.  If one is
interested in operator insertions not discussed here, it will be
necessary to extend this work by computing the renormalizations of those
insertions. 

We will present numerical values of renormalization constants in some
particular theories, and give step by step instructions for building
$O(a)$ improved Lagrangians, in Appendix B.

One final comment is in order.  As we have written them, the corrections
$Z_m$, $Z_g$, $\delta \lambda$ etc which are needed for this $O(a)$
improvement are to be computed with the $O(a)$ improved parameters
$\lambda$, $a$, etc, not with $\lambda_n$, $a_n$, etc.  If one only
knows the ``naive'' values, this might
require an iterative or ``bootstrap'' type calculation.  However, the
difference between using the improved and unimproved parameters in the
calculation is formally $O(a^2)$, so one should not worry too much which
is used--except for computing the one loop $O(1/a)$ corrections to
masses and order parameters, as already discussed.

\section{What to do in SU($N$)$\times$SU($M$)}
\label{extension}

Sometimes it is interesting to treat a case where the gauge group is not
simple.  Here we will restrict our attention to SU($N$)$\times$SU($M$), 
where each
scalar field is a singlet under one or the other group, and we intend to
be illustrative but not exhaustive, so we will consider the case where
an SU($N$) group has both an adjoint and a fundamental scalar (it is
easy to delete one of them) and there is one more scalar which
transforms trivially under SU($N$) but perhaps nontrivially under the other
group, and has self-interactions.  Calling the added scalar $s$, and
writing $s^2$ to mean $2s^\dagger s$ if it is fundamental and $2{\rm Tr}
s^2$ if it is adjoint, the new interactions between the $s$ field and
the scalar fields $\phi$ and $\Phi$ which can appear in the Lagrangian
are
\begin{equation}
\frac{h_{s1}}{2} s^2 \phi^\dagger \phi + \frac{h_{s2}}{2} s^2 {\rm Tr}
	\Phi^2 \, .
\end{equation}
If $h_{s1} = 0 = h_{s2}$ then the theory decouples into two
noninteracting parts, the SU($N$) part and the part which contains the
scalar $s$.  We assume that one can compute the corrections in this
limit by using the last section, and we will compute the new corrections
which must be added at nonzero $h_{s1}$ or $h_{s2}$.  We will compute
the corrections to the couplings of the SU($N$) sector and $h_{s1}$,
$h_{s2}$ with the minimum information about the scalar $s$.  To compute
corrections to the sector containing the $s$ field, just think of it as
the SU($N$) sector and use these same results.

We need only a little information about the scalar $s$ and its (gauge
and self) couplings to compute the corrections to the SU($N$) sector.  
What we need to know is the number of degrees of freedom $s$ contains, 
$N_{\rm
DOF}$ (in SU($M$) this is $2M$ for a complex fundamental field and
$M^2-1$ for a real adjoint field), 
and its one loop mass renormalization at $h_{s1} = h_{s2} = 0$,
\begin{equation}
\delta m_s^2 ({\rm one \; loop \, , \; zero \; }h_{s1} , h_{s2} )
	= - \frac{\Sigma}{4 \pi a} \left( C_1 + C_2 \right) \, ,
\end{equation}
where $C_1$ is the contribution from all scalar couplings and $C_2$ is
the contribution from any gauge couplings it participates in.  They have
the same units as a coupling constant.

The one loop mass corrections due to the new particle are 
\begin{eqnarray}
\delta m_\phi^2 & += & - \frac{\Sigma}{4 \pi a} 
	\frac{N_{\rm DOF}}{2} h_{s1} \, , \\
\delta m_\Phi^2 & += & - \frac{\Sigma}{4 \pi a} 
	\frac{N_{\rm DOF}}{2} h_{s2} \, .
\end{eqnarray}
there are no new one loop additive operator insertion corrections.

The one loop $O(a)$ corrections to scalar self-couplings are increased by
\begin{eqnarray}
\delta \lambda & += & \frac{\xi a}{4 \pi} \frac{N_{\rm DOF}}{4}
	h_{s1}^2 \, , \\
\delta \lambda_1 & += & \frac{\xi a}{4 \pi} \frac{N_{\rm DOF}}{4}
	h_{s2}^2 \, , \\
\delta \lambda_2 & += & 0 \, , \\
\delta h_1 & += & \frac{\xi a}{4 \pi} \frac{N_{\rm DOF}}{2}
	h_{s1} h_{s2} \, , \\
\delta h_2 & += & 0 \, , \\
\label{deltahs1}
\delta h_{s1} & = & \frac{a}{4 \pi} \Bigg\{ \left( C_1 h_{s1} + 
	2(N+1) \lambda h_{s1} + \frac{N^2-1}{2} h_1 h_{s2} + 2 h_{s1}^2 
	\right) \xi - \nonumber \\ & & \qquad
	- \left( \frac{N^2-1}{12 N} g^2 + \frac{C_2}{12}
	\right) \left( 18 \xi + \Sigma \right) h_{s1} \Bigg\} \, , \\
\label{deltahs2}
\delta h_{s2} & = & \frac{a}{4 \pi} \Bigg\{ \left( C_1 h_{s2} + 
	(N^2+1) \lambda_1 h_{s2} + \frac{2N^2-3}{N} \lambda_2 h_{s2}
	+ N h_1 h_{s1} + 2 h_{s2}^2 
	\right) \xi - \nonumber \\ & & \qquad
	- \left( \frac{N}{6} g^2 + \frac{C_2}{12}
	\right) \left( 18 \xi + \Sigma \right) h_{s2} \Bigg\} \, .
\end{eqnarray}
There are no corrections to the gauge field renormalization $Z_g$ at
this order.  Such corrections actually start at $O(a^3)$.

Moreover, we should now think of $Z_{OP}$ as a $3 \times 3$ matrix,
which at $h_{s1} = 0 = h_{s2}$ is nonzero in the upper $2 \times 2$
block, where it equals $Z_{OP}$ of the last chapter, and the lower right
component, where it is $Z_{OP}$ for $s$.  There are no new contributions
in these blocks.  The terms we should compute are
\begin{equation}
Z_{OP} - {\bf 1}  =  \frac{\xi}{4 \pi a} \left[ \begin{array}{ccc}
	{\rm same} & {\rm same} & N h_{s1} \\
	{\rm same} & {\rm same} & \frac{N^2-1}{2} h_{s2} \\
	\frac{N_{\rm DOF}}{2} h_{s1} & \frac{N_{\rm DOF}}{2} h_{s2} &
	{\rm same} \\ \end{array} \right] \, ,
\label{newZop}
\end{equation}
where ``same'' means the same value as at $h_{s1} = 0 = h_{s2}$.
At one loop there are no corrections to wave function 
renormalizations due to $s$.  These will first appear at $O(a^2)$.

What remains is to compute two loop mass and operator insertion additive
corrections.  The only new correction to the operator insertion
conterterm is that the new value of $Z_{OP}$, Eq. (\ref{newZop}), 
should be used in Eq.
(\ref{phicounter}).  The new corrections to the mass counterterm, aside
from using the new values for $\delta \lambda$ etc. in Eqs.
(\ref{newmcounter1},\ref{newmcounter2}), are
\begin{eqnarray}
\delta m_\phi^2 & += & - \frac{N_{\rm DOF}}{2} \left[ 
	\frac{\Sigma}{4 \pi a}
	\delta h_{s1} + \frac{h_{s1}}{16 \pi^2}
	\left( ( C_2 -h_{s1} ) \left( \ln \frac{6}{a \mu}
	+ \zeta \right) + C_2 \left( \frac{\Sigma^2}{4} - \delta 
	\right) \right) \right] \, , \\
\delta m_\Phi^2 & += & - \frac{N_{\rm DOF}}{2} \left[ 
	\frac{\Sigma}{4 \pi a}
	\delta h_{s2} + \frac{h_{s2}}{16 \pi^2}
	\left( ( C_2 -h_{s2} ) \left( \ln \frac{6}{a \mu}
	+ \zeta \right) + C_2 \left( \frac{\Sigma^2}{4} - \delta 
	\right) \right) \right] \, .
\end{eqnarray}

If there are more than one scalar which are SU($N$) singlets, the
corrections are just the sum of the corrections from each particle,
except for cross terms in the renormalization of the $h_s$ type
interactions, which can be figured out from Eqs.
(\ref{deltahs1},\ref{deltahs2}).  Things become more complicated 
at higher loop order, however.  If there are 3 or more
gauge groups, each with scalars, then the combinatorics of the one loop
scalar coupling corrections get messy and what we show here may not be
sufficiently general.  But we know of no physically interesting theories
in this class.  Also, if there are scalars which transform nontrivially
under more than one gauge group, the situation becomes considerably more
complicated.  For instance, the one 
loop self-coupling and two loop mass counterterms
then contain $g_1^2 g_2^2$ type terms.  The particularly simple 
case of SU(2)$\times$U(1) is treated in \cite{Oapaper,LaineRajantie}.  
We will not attempt anything more general here.

\section{Conclusion}
\label{conclusion}

We have discussed why it is possible to write $O(a)$ corrected actions
for 3 dimensional lattice Yang-Mills Higgs theories, and we have
extended the previous results for SU(2)$\times$U(1) to SU($N$) with a
fundamental and/or an adjoint Higgs field.  We have also shown how the
corrections can often be applied to uncorrected data ``after the fact.''
Generally the result is that the actual value of 3-D scalar
self-coupling used is lower than the naive coefficient put into the
lattice calculation, and actual jumps in the order parameters 
are smaller than the uncorrected data imply.

The limitation of the work here is that we have not been able to compute
the $O(a)$ additive corrections to the Higgs field masses or the
$\phi^2$ operator insertions.  Computing these is possible in principle
but requires doing a lot of very nasty 3 loop diagrams.  We have
argued that for many purposes these are the least important $O(a)$
errors, and that the knowledge of the relation between the 3 dimensional
and physical, thermal 4 dimensional theories is in any case weaker here.
The corrections for the other coefficients are definitely valuable for
numerical efforts to study hot gauge theories nonperturbatively,
allowing better precision with less numerical effort.

A final issue we should address is:  how small should $a$ be made so
that we don't have to worry about the remaining $O(a^2)$ errors?  Of
course, the answer depends on what one is doing.  For instance, if one
is studying the strength of a phase transition where it is strong, say
at very small $\lambda$ or $\lambda_1$, then the natural scale of
physics involved becomes shorter than $1/g^2$ and nonrenormalizable
operators may be a problem.  One can estimate the importance of these
operators by including them in a perturbative calculation of the
effective potential, as was done in \cite{Oapaper}.  The conclusion is
that nonrenormalizable derivative terms are less important than the
$O(a)$ corrections $\delta \lambda$ to the scalar self-couplings, roughly
by a factor of $g a \sqrt{\phi^\dagger \phi}$.  If neglecting the $O(a)$
scalar self-coupling corrections would make a significant difference 
and  $g a \sqrt{\phi^\dagger \phi}$ is not very small, 
then $a$ is not small enough.

Away from the case where the phase transition is strong, the two loop,
$O(a^2)$ corrections to the Lagrangian parameters may also be important.
We expect that the largest of these will be ``tadpole'' type corrections
to the gauge field normalization, higher loop analogs of 
the first term in Eq. (\ref{gaugerenorm}).  This term arises from a
nonrenormalizable operator in the action of form ${\rm Tr} F_{ij}^4$.
It turns out to dominate $Z_g^{-1} - 1$, which in turn is the largest of
the $O(a)$ corrections.  We might expect that ``tadpole'' type
corrections, from multiple appearances of this term and from the term of
form ${\rm Tr} F_{ij}^6$, give the most important contributions to
$Z_g^{-1} - 1$ at two loops, and that this is again the largest $O(a^2)$
correction.  (This would follow a fairly general pattern of behavior for
lattice gauge theories.)  Fortunately, it is not too hard to compute
this limited class of 2 loop diagrams; we do so in Appendix A.  The
conclusion is that the two loop correction to $Z_g^{-1} - 1$ is about
the same size as the square of the one loop correction.  So if the one
loop correction is $10 \%$, then the two loop correction is $1 \%$.  If
nonrenormalizable operators are not a problem, then the one loop
improvement is likely to be sufficient as long as its coefficients make
corrections in the $10 \%$ range.

It may not always be possible to reduce $O(a^2)$ errors to the level
desired, and sometimes it is also important to know mass parameters with
high precision.  In these cases it may still be necessary to perform 
an extrapolation over data at several values of $a$.  We expect that 
the $O(a)$ corrections will still be useful in this case, because
they should make the extrapolation much better behaved.

\medskip

\centerline{Acknowledgements}

I am indebted to Keijo Kajantie for suggesting this topic and for
correspondence, and to Mikko Laine, Arttu Rajantie, Kari Rummukainen,
and particularly Misha Shaposhnikov for correspondence or discussions.

\appendix

\section{Two loop tadpole type diagrams}

In this appendix we compute two loop ``tadpole'' corrections
to the gauge field self-energy.  By ``tadpole'' corrections we mean
corrections emerging from a certain restricted subset of the interaction
terms generated by the gauge field term in the action.  Namely, we
approximate that $P_{ij} = \exp ( i g a^2 T^a F^a_{ij} )$, where 
$F_{ij}$ is linear in the gauge field, ie in Fourier space it is
$F^a_{ij}(k) = k_i A^a_j(k) - k_j A^a_i(k)$.  This amounts to neglecting
appearances of $f_{abc}$ in favor of $\delta_{ab}$ and $d_{abc}$.

The terms in the action which emerge from this approximation are
\begin{equation}
a^3 \sum_x \left( \sum_{i \neq j} \frac{1}{2 Z_g} {\rm Tr} F_{ij}^2
	- \sum_{i \neq j} \frac{g^2 a^4}{24 Z_g} {\rm Tr} F_{ij}^4
	+ \sum_{i \neq j} \frac{g^4 a^8}{720 Z_g} {\rm Tr} F_{ij}^6 
	- \ldots \right) \, .
\end{equation}
These give the gauge field propagator and a tower of interaction terms.
The term with $2n$ powers of $F_{ij}$
gives a vertex which attaches to $2n$ lines, with momenta we will label
as $p_1, \: p_2, \ldots$, polarization indicies we label as $i_1, \: i_2,
\ldots$, and group indicies we label as $a, \: b, \ldots$.  The Feynman
rule for the vertex is that it contributes
\begin{eqnarray}
& & g^{2n-2} a^{4n-4} (-1)^{n} \sum_{jk} ( i p_{1,j} \delta_{i_1,k} 
	- i p_{1,k} \delta_{i_1,j} )( i p_{2,j} \delta_{i_2,k} 
	- i p_{2,k} \delta_{i_2,j} ) \ldots 
	\delta ( p_1 + p_2 + \ldots )\times \nonumber \\
& & \qquad \times \frac{1}{(2n)!} \left( 
	{\rm Tr} ( T^a T^b \ldots ) + {\rm all \; permutations \; of \;}
	a,b,\ldots \right) \, .
\end{eqnarray}

\begin{figure}[b]
\centerline{\mbox{\psfig{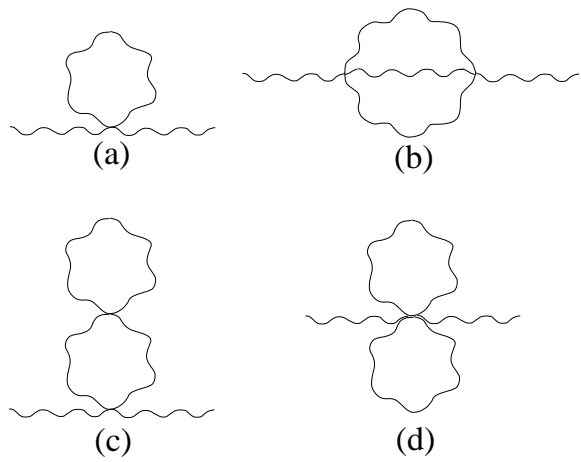}}}
\caption{\label{selffig} Tadpole type diagrams which correct the gauge
field self-energy at one and two loops.}
\end{figure}

Though the nonrenormalizable terms here have several powers of $a$, they
lead to diagrams with strong power law UV divergences and are important
to the radiative corrections.  The first of these is responsible,
through diagram $(a)$ in Figure \ref{selffig}, for the first term in
Eq. (\ref{gaugerenorm}).

The first natural question is, is this approximation any good?  To
answer that, compare the numerical value of the first term in Eq.
(\ref{gaugerenorm}) to the sum of the others.  Even with both scalars
present, in SU(2) the first term is larger than the others combined by a
factor of 3.  At large $N$ the domination is by a factor of 6.  Without
scalars the domination is twice as large.  So it appears that
``tadpole'' type terms do give the dominant contributions.  We should
also comment that the contributions from these diagrams are gauge
invariant and transverse diagram by diagram.

Now let us compute the two loop ``tadpole'' type terms.  There are
insertions of the $O(a)$ counterterm in the propagator and the vertex of
diagram $(a)$, but these cancel.  There are also diagrams $(b)$, $(c)$,
and $(d)$, which give contributions to $Z_g^{-1} - 1$ equal to
\begin{eqnarray}
(b) & \rightarrow & \frac{g^4 a^2 \pi^2}{16 \pi^2} 
	\left( \frac{N^4 - 6N^2 + 10} {6 N^2} \right) 
	\left( \frac{2 \times 0.3015814}{3} \right)  \, , \\
(c) & \rightarrow & \frac{g^4 a^2 \pi^2}{16 \pi^2} \left(
	\frac{(2N^2 - 3)^2}
	{9 N^2}\right) \left( \frac{4}{9} \right) \, , \\
(d) & \rightarrow & - \frac{g^4 a^2 \pi^2}{16 \pi^2} \left( 
	\frac{2 ( 5N^4 - 15N^2 - 1 ) }{15 N^2} \right)
	\left( \frac{1}{9} \right) \, .
\end{eqnarray}
We have written each contribution as a common ``loop counting'' type
term, a group theoretic term (with powers of $g^2/4$ taken out into the
leading expression), and the product of the symmetry factor and
momentum integration.  Diagram $(c)$ gives precisely the square of the
one loop contribution.
For every value of $N$, the sum of these 2 loop diagrams 
is less than but on order the square of the one loop correction.

The numerical constant, $0.3015814$, is the integral
\begin{equation}
\int_{ [ -\pi , \pi ]^9} \frac{ d^3 k d^3 l d^3 m}{(2 \pi)^9}
	\left(
	\frac{\tilde{k}_1^2 + \tilde{k}_2^2}{\tilde{k}^2}
	\right)
	\left(
	\frac{\tilde{l}_1^2 + \tilde{l}_2^2}{\tilde{l}^2}
	\right)
	\left(
	\frac{\tilde{m}_1^2 + \tilde{m}_2^2}{\tilde{m}^2}
	\right)
	(2 \pi)^3 \delta^3 ( k + l + m ) \, ,
\end{equation}
where $\tilde{k}_i = 2 \sin ( k_i / 2 )$ and $\tilde{k}^2 = \sum_i
\tilde{k}_i^2$.  This integral, like all the integrals arising from the
tadpole terms, is completely infrared safe and has no continuum anolog.

\section{Some particular examples}

In this appendix we will plug in numerical values and give step by step
instructions for building $O(a)$ improved Lagrangians for the specific
theories of most interest, namely SU(2) plus fundamental Higgs, SU(3) and
SU(5) plus adjoint Higgs, and SU(3)$\times$SU(2), each with a
fundamental Higgs.  The latter is a model for studying the phase
transition in the minimal supersymmetric standard model with a light
stop squark, which makes including the SU(3) color sector necessary.

In each case the first step is to write down the desired values for $g^2
a / 4$, $x_i = \lambda_i / g^2$, and $y_i = m_i^2 / g^4$, which one
wants the simulation to correspond to.  Then one computes a number of
counterterms and uses them to find the ``naive'' values $g^2 a_n/4$,
$x_{i,n}$, $y_{i,n}$ corresponding to the desired improved values.  One
constructs the lattice Lagrangian using these, computing two loop mass
squared and operator counterterms $\delta m^2$ and $\delta \langle \phi^2
\rangle$ exactly as one would in ignorance of the $O(a)$ improvement
scheme, using expressions in \cite{FKRS,Laine,LaineRajantie}.  Then one
applies $O(a)$ multiplicative corrections to ``naive'' measured values
of $(\langle \phi^2 \rangle + \delta \langle \phi^2 \rangle)_{\rm naive}$
and the surface tension $\sigma_{\rm naive}$.

\subsection{SU(2) + fundamental Higgs}

We begin with SU(2) plus fundamental Higgs theory.  Writing $x = \lambda
/ g^2$ and $y = m^2 / g^4$, we choose desired values for $g^2 a / 4$,
$x$, and $y$, and then compute the counterterms
\begin{eqnarray}
Z_g^{-1} - 1 & = & 0.6674 \frac{g^2 a}{4} \, , \\
\delta x & = & \left( 0.01825 - 0.4717 x + 0.5839 x^2 \right) 
	\frac{g^2 a}{4} \, , \\
Z_m - 1 & = & \left( - 0.2358 + 0.2919 x \right) 
	\frac{g^2 a}{4} \, .
\end{eqnarray}

Now compute the ``naive'' values 
\begin{eqnarray}
g^2 a_n & = & Z_g g^2 a \, , \\
x_n & = & Z_g^{-1} ( x + \delta x ) \, , \\
y_n & = & Z_g^{-2} Z_m y \, ,
\end{eqnarray}
and construct a lattice lagrangian which, according to naive lattice to
continuum relations, should model the continuum theory with these
``naive'' values.  Compute the counterterms $\delta y_n$ and 
$(\delta \langle \phi^{\dagger} \phi \rangle)_{n}$ using the two loop
expressions in \cite{Laine},
and using $a_n$, $x_n$.  Do not include the corrections, Eqs.
(\ref{newmcounter1},\ref{newmcounter2},\ref{phicounter}).  
(Alternately, compute them using $g^2a$ and $x$, and include
Eqs. (\ref{newmcounter1},\ref{newmcounter2},\ref{phicounter}), but
then multiply by $Z_g^{-2}Z_m$.  The two approaches differ at $O(g^2a)$
and it is not clear without the 3 loop calculation which will result in
smaller $O(a)$ additive errors.)

This theory will in fact represent the continuum theory with lattice
spacing $a$, scalar self-coupling $\lambda = x g^2$, and Higgs mass
$m_\phi^2 = y g^4$, up to $O(a^2)$ errors 
(and an $O(a)$ additive shift to $y$).  
The $O(a)$ corrected value for the order parameter is
\begin{equation}
\langle \phi^\dagger \phi \rangle_{\rm corrected} = Z_g Z_m
	\left( \langle \phi^\dagger \phi \rangle_{n,{\rm meas}}
	+ (\delta \langle \phi^\dagger \phi \rangle)_n \right) \, ,
\end{equation}
and the surface tension is
\begin{equation}
\sigma_{\rm corrected} = Z_g^2 \sigma_{\rm naive} \, .
\end{equation}
Note that in 3-D units the surface tension is just an inverse area.  To
convert to 4-D units one multiplies by $T$.  Physical lengths such as
correlation lengths should be corrected by multiplying by $Z_g^{-1}$.

\subsection{SU(3) + adjoint Higgs}

Suppose we want an improved lattice action for SU(3) plus adjoint Higgs
theory at a desired value of $g^2 a /4$, $x = (\lambda_1 + \lambda_2 /
2) / g^2$, and $y = m_\Phi^2 / g^4$.  We compute
\begin{eqnarray}
Z_g^{-1} - 1 & = & 1.3299 \frac{g^2 a}{4} \, , \\
\delta x & = & \left( 0.21895 - 1.8867 x + 0.7785 x^2 \right) 
	\frac{g^2 a}{4} \, , \\
Z_m - 1 & = & \left( - 0.9434 + 0.4866 x \right) 
	\frac{g^2 a}{4} \, .
\end{eqnarray}
Note that the pure gauge (no powers of $x$) corrections are much larger
here than in SU(2) plus fundamental Higgs theory, 
particularly for the self-coupling.  This means we need a smaller value
of $g^2 a/4$ to fight down lattice artifacts, especially if we are
exploring small $x$.

From this point the steps are the same as in SU(2); write down a lattice
action which according to ``naive'' lattice to continuum matching
corresponds to the continuum theory at the ``naive'' values
\begin{eqnarray}
g^2 a_n & = & A_g g^2 a \, , \\
x_n & = & Z_g^{-1} ( x + \delta x ) \, , \\
y_n & = & Z_g^{-2} Z_m y \, ,
\end{eqnarray}
and compute the counterterms for the mass squared and operator
insertion from the equations in \cite{LaineRajantie}.
Take data, and correct the ${\rm Tr} \Phi^2$ order parameter by
multiplying by $Z_g Z_m$ and the surface tension by multiplying by
$Z_g^2$.

\subsection{SU(5) + adjoint Higgs}

This case differs from SU(3) plus adjoint Higgs theory only in that
there are now two scalar self-couplings $x_1 = \lambda_1/g^2$ and $x_2 =
\lambda_2 / g^2$, and that there are different numerical values for the
counterterms, 
\begin{eqnarray}
Z_g^{-1} - 1 & = & 2.454 \frac{g^2 a}{4} \, , \\
\delta x_1 & = & \left( 0.1460 - 3.145 x_1 + 1.557 x_1^2 + 
	0.915 x_1 x_2 + 0.163 x_2^2 \right) 
	\frac{g^2 a}{4} \, , \\
\delta x_2 & = & \left( 0.2433 - 3.145 x_2 +
	0.584 x_1 x_2 + 0.311 x_2^2 \right) 
	\frac{g^2 a}{4} \, , \\
Z_m - 1 & = & \left( - 1.572 + 1.265 x_1 + 0.457 x_2 \right) 
	\frac{g^2 a}{4} \, .
\end{eqnarray}
From here on the procedure follows the SU(3) plus adjoint Higgs case.
Note that $g^2a/4$ must be even smaller for SU(5) than for SU(3), which
is not too surprising.

\subsection{SU(3)$\times$SU(2) + fundamental Higgs in each}

This theory could be phenomenologically interesting if the left handed
stop squark is light enough that a perturbative treatment becomes
unreliable.  It holds a potentially rich phenomenology, allowing a
double phase transition and an exotic color broken symmetric 
phase \cite{Lainestop}.  It also nicely illustrates what to do when
there are more than one gauge group, and it presents a heirarchy
problem which makes its numerical study without $O(a)$ corrections
almost impossible, and its study even with them rather tenuous, for the
physical value of $g_s^2 / g_w^2$.

We will use the notation of Laine and Rajantie.  The continuum
Lagrangian is
\begin{eqnarray}
{\cal L} & = & \frac{1}{2} {\rm Tr} G_{ij}^2 + \frac{1}{2} {\rm Tr} 
	F_{ij}^2 
	+ (D_i U)^\dagger (D_i U)
	+ (D_i H)^\dagger(D_i H) + \nonumber \\ & & 
	+ m_U^2 U^\dagger U + m_H^2 H^\dagger H +
	\lambda_U (U^\dagger U)^2 + \lambda_H (H^\dagger H)^2 +
	\gamma U^\dagger U H^\dagger H \, ,
\end{eqnarray}
where $U$, the squark field, is in the fundamental representation of
SU(3) (the field $G$, with coupling constant $g_s$) and $H$, the light
Higgs field, is in the fundamental representation of SU(2) (the field
$F$, with coupling constant $g_w$).

The theory is characterized by one dimensionful number $g_s^2$ and 6
dimensionless parameters, $x_1 = \lambda_U/g_s^2$, $x_2 = \lambda_H /
g_s^2$, $x_3 = \gamma/g_s^2$, $y_1 = m_U^2 / g_s^4$, $y_2 = m_H^2 /
g_s^4$, and $z = g_w^2 / g_s^2$.  Note that we express all of the masses
and couplings in terms of $g_s^2$.  To put the theory on the lattice, we
first choose a desired value for 
each dimensionless number and a value for the
lattice spacing in terms of the one dimensionful number, that is we must
choose $g_s^2 a / 4$.  The $O(a)$ improvement is carried out by first
computing the counterterms
\begin{eqnarray}
Z_s^{-1} - 1 & = & 1.2619 \frac{g_s^2 a}{4} \, , \\
Z_w^{-1} - 1 & = & (0.6674) z \frac{g_s^2 a}{4} \, , \\
\delta x_1 & = & \left( 0.03514 - 0.8386 x_1 + 0.6812 x_1^2 
	+ 0.0487 x_3^2 \right) \frac{g_s^2 a}{4} \, , \\
\delta x_2 & = & \left( 0.01825 z^2 - 0.4717 z x_2 + 0.5839 x_2^2
	+ 0.0730 x_3^2 \right) \frac{g_s^2 a}{4} \, , \\
\delta x_3 & = & \left( - 0.4193 x_3 - 0.2358 z x_3 + 0.3893 x_1 x_3
	+ 0.2919 x_2 x_3 \right) \frac{g_s^2 a}{4} \, , \\
Z_m - {\bf 1} & = & \frac{g_s^2 a}{4} \left[ \begin{array}{cc}
	-0.4193 + 0.3893 x_1 & 0.1460 x_3 \\
	0.0973 x_3 & -0.2358 z + 0.2919 x_2 \\ \end{array} \right] \, ,
\end{eqnarray}
where we choose the convention that $U^\dagger U$ goes in the upper and
$H^\dagger H$ in the lower row in expressions where $Z_m$ acts on
columns.

Now we implement the $O(a)$ improvement by defining a set of ``naive''
paramters, 
\begin{eqnarray}
a_n & = & Z_s a \, , \\
z_n & = & \frac{Z_w}{Z_s} z \, , \\
x_{1,n} & = & Z_s ( x_1 + \delta x_1 ) \, , \\
x_{2,n} & = & Z_s ( x_2 + \delta x_2 ) \, , \\
x_{3,n} & = & Z_s ( x_3 + \delta x_3 ) \, , \\
\left[ \begin{array}{c} y_{1,n} \\ y_{2,n} \\ \end{array} \right]
	 & = & Z_s^{-2} Z_m \left[ \begin{array}{c} y_1 \\ y_2 \\
	\end{array} \right] \, .
\end{eqnarray}
We then use a ``naive'' continuum to lattice relation to convert these
into a lattice action, computing the mass and operator insertion
conterterms from the expressions in \cite{LaineRajantie}.  Finally, 
improved values for jumps in order parameters are computed by
multiplying the unimproved values by the matrix $Z_s Z_m$, the
surface tension is improved by multiplying by $Z_s^2$, and correlation
lengths are rescaled by $Z_s^{-1}$.

One point is in order, though.  For $T \simeq 100$GeV, the values of the
coupling constants are around\footnote{To 
some readers the value for $\alpha_s$ 
may seem low, since the value at the $Z$ pole is about
0.118.  But the 3-D theory value is roughly the value at the
$\overline{\rm MS}$ renormalization point $\mu = 7 T$, and $\alpha_s$
runs quite a bit in between.} 
$g_s^2/4 \pi \simeq 0.087 T$ and $g_w^2/4 \pi \simeq 0.032T$.  
Hence, $z \simeq .368 \sim 1/3$, so the ``natural scale'' of the
strong and weak sectors differ by a factor of 3.  To contain $O(a^2)$
errors in the strong sector, we need $g_s^2 a / 4$ to be fairly
small--for instance, we probably want $Z^{-1}_s - 1$ to be on order 0.1.
However, this forces a value of $(g_w^2 a / 4) < 1/30$, which is a very
fine lattice.  Yet we must make the lattice volume large enough to 
contain the rather long correlation lengths of the SU(2) sector, which
puts some strain on numerical practicality.  Hence there may
realistically be something of a heirarchy problem involved in studying
this system at the physical values for the couplings.

\end{document}